\documentclass[fleqn,usenatbib]{mnras}

\usepackage{newtxtext,newtxmath}
\usepackage[T1]{fontenc}

\DeclareRobustCommand{\VAN}[3]{#2}
\let\VANthebibliography\thebibliography
\def\thebibliography{\DeclareRobustCommand{\VAN}[3]{##3}\VANthebibliography}

\usepackage{graphicx}	% Including figure files
\usepackage{amsmath}	% Advanced maths commands
\usepackage{multirow}
\usepackage[utf8]{inputenc}
\usepackage[capitalise, nameinlink]{cleveref}
\usepackage{csquotes}
\usepackage{gensymb}

\crefname{section}{§}{§§}
\Crefname{section}{§}{§§}

\title[Comptonization model Type-B QPOs in GX 339$-$4]{Dual-Corona Comptonization model for the Type-B Quasi-Periodic Oscillations in GX 339$-$4}

\author[V. Peirano et al.]{
Valentina Peirano,$^{1}$\thanks{E-mail: v.peirano@astro.rug.nl}
Mariano M\'endez,$^{1}$
Federico Garc\'ia$^{2}$
and Tomaso Belloni$^{3}$
\\
$^{1}$Kapteyn Astronomical Institute, University of Groningen, P.O. BOX 800, 9700 AV Groningen, The Netherlands\\
$^{2}$Instituto Argentino de Radioastronomía (CCT La Plata, CONICET; CICPBA; UNLP), C.C.5, (1894) Villa Elisa, Buenos Aires, Argentina\\
$^{3}$INAF-Osservatorio Astronomico di Brera, via E. Bianchi 46, I-23807 Merate, Italy
}

\date{Accepted XXX. Received YYY; in original form ZZZ}

\pubyear{2022}

\begin{document}
\label{firstpage}
\pagerange{\pageref{firstpage}--\pageref{lastpage}}
\maketitle

\begin{abstract}
Characterising the fast variability in black-hole low-mass X-ray binaries (BHXBs) can help us understand the geometrical and physical nature of the innermost regions of these sources. Particularly, type-B quasi-periodic oscillations (QPOs), observed in BHXBs during the soft-intermediate state (SIMS) of an outburst, are believed to be connected to the ejection of a relativistic jet. The X-ray spectrum of a source in the SIMS is characterised by a dominant soft blackbody-like component – associated with the accretion disc – and a hard component – associated with a Comptonizing region or corona. Strong type-B QPOs were observed by \textit{NICER} and \textit{AstroSat} in GX 339$-$4 during its 2021 outburst. We find that the fractional rms spectrum of the QPO remains constant at $\sim$1 per cent for energies below $\sim$1.8 keV and then increases with increasing energy up to $\sim$17 per cent at 20-30 keV. We also find that the lag spectrum is \textit{"U-shaped"}, decreasing from $\sim$1.2 rad at 0.7 keV to 0 rad at $\sim$3.5 keV, and increasing again at higher energies up to $\sim$0.6 rad at 20-30 keV. Using a recently developed time-dependent Comptonization model, we fit simultaneously the fractional rms and lag spectra of the QPO and the time-averaged energy spectrum of GX 339$-$4 to constrain the physical parameters of the region responsible for the variability we observe. We suggest that the radiative properties of the type-B QPOs observed in GX 339$-$4 can be explained by two physically-connected Comptonizing regions that interact with the accretion disc via a feedback loop of X-ray photons.
\end{abstract}


%%%%%%%%%%%%%%%%%%%%%%%%%%%%%%%%%%%%%%%%%%%%%%%%%%

%%%%%%%%%%%%%%%%% BODY OF PAPER %%%%%%%%%%%%%%%%%%

\section{Introduction}
Our understanding of the nature of the fast variability in the X-ray emission observed in black-hole (BH) low-mass X-ray binaries (LMXBs) and the analysis techniques used to study this variability are continuously changing. These variability features, better known as quasi-periodic oscillations (QPOs), are believed to originate in the innermost parts of the BH systems, precisely in the regions where the most extreme physical phenomena take place.

QPOs appear in the power density spectra (PDS) of BH LMXBs over a wide range of frequencies, from 0.1 to 30 Hz \citep[see e.g.][]{belloniTransientBlackHole2016,ingramReviewQuasiperiodicOscillations2019}. Particularly, low-frequency QPOs are classified into three categories – type-A, B and C – according to their strength in the PDS, central frequency ($\nu_0$), full-width at half-maximum (FWHM), the properties of the accompanying broadband noise (BBN) present in the PDS and the spectral state of the source \citep[see e.g.][]{casellaABCLowFrequencyQuasiperiodic2005}. Specifically, type-B QPOs are observed while the source undergoes a transition between the low-hard and high-soft state: the soft-intermediate state (SIMS) \citep[][]{belloniStatesTransitionsBlack2010}. During this spectral state, the energy spectrum of the BH is dominated by the emission of an optically thick, geometrically thin accretion disc \citep[][]{shakuraBlackHolesBinary1973} with contribution of a hard component, probably originated in a Comptonizing region or corona, where Compton up-scattering of the photons from the disc occurs \citep[][]{thorneCygnusX1Interpretation1975,sunyaevHardXraySpectrum1979}.

Type-B QPOs have central frequencies from 4 to 6 Hz, a strong and narrow peak, with a quality factor $Q = \nu_0/\mathrm{FWHM} > 6$ and fractional rms amplitude of around 4\% \citep[][]{casellaABCLowFrequencyQuasiperiodic2005}. Other variability features in the PDS, such as red noise that increases at low frequencies, and the second harmonic and sub-harmonic peaks, generally appear weaker than the fundamental/type-B QPO \citep[see e.g.][]{casellaStudyLowfrequencyQuasiperiodic2004}. The presence of type-B QPOs in the PDS of a BH LMXB is considered an unequivocal indication that the source is in the SIMS \citep[][]{belloniTransientBlackHole2016}.

Rapid state transitions sometimes occur in BH systems, where type-B QPOs appear and disappear in the PDS within timescales of just a few seconds \citep[e.g.][]{miyamotoXrayVariabilityGX1991,takizawaSpectralTemporalVariability1997,casellaStudyLowfrequencyQuasiperiodic2004,belloniEvolutionTimingProperties2005,zhangNICERUncoversTransient2021,bogensbergerUnderlyingClockExtreme2020}. Relativistic jet ejections have been observed during these state transitions, which suggests that there is a relation between the jet and the mechanism responsible for the QPOs \citep[e.g.][]{fenderJetsBlackHole2009,russellDiskJetCoupling20172019,homanRapidChangeXRay2020}. Particularly for GX 339$-$4, \citet{kylafisQuantitativeExplanationTypeB2020} proposed a model where Comptonization takes place in a jet that precesses at the QPO frequency, producing the variability in the photon-number spectral index of the energy-spectrum found by \citet{stevensPhaseresolvedSpectroscopyType2016}. To explain the radiative properties of the variability in the emission of LMXBs, other models consider that Comptonization occurs in a lamp-post corona, situated above the BH and the accretion disc \citep[see e.g.][]{ingramPublicRelativisticTransfer2019,mastroserioModellingCorrelatedVariability2021}, that interacts with the disc through reverberation producing the lags we observe \citep[see e.g.][]{uttleyCausalConnectionDisc2011,uttleyXRayReverberationAccreting2014,demarcoTracingReverberationLag2015,wangNICERReverberationMachine2022}. Despite the fact that these models can characterise the time-lags in BH systems, they do not describe the fractional rms amplitude of the QPOs.

To explain simultaneously both the energy-dependent rms amplitude and lags of high-frequency QPOs in neutron-star systems, and based on the concepts introduced by \citet{leeComptonizationQPOOrigins1998}, \citet{leeComptonUpscatteringModel2001} and \citet{kumarEnergyDependentTime2014}, \citet{karpouzasComptonizingMediumNeutron2020} developed a Comptonization model that considers the presence of a feedback loop between the hard and soft components of the LMXB. In this feedback loop, a fraction of Comptonized hard photons emitted by the corona impinge back onto the soft-photon source – either the accretion disc or the surface of the compact object – and are subsequently emitted at lower energies and later times. Recently, \citet{bellavitaVKompthVariableComptonization2022} adapted the model of \citet{karpouzasComptonizingMediumNeutron2020} to describe the properties of QPOs in BH systems by considering that the soft-photon source is the accretion disc instead of the surface of the compact object. The model of \citet{bellavitaVKompthVariableComptonization2022}, \textsc{vkompth}, has been successfully used to explain the low-frequency QPOs in BH LMXBs like MAXI J1348$-$630 \citep[][]{garciaTwocomponentComptonizationModel2021} and GRS 1915$+$105 \citep[][]{karpouzasVariableCoronaGRS2021,mendezCouplingAccretingCorona2022,garciaEvolvingPropertiesCorona2022} and MAXI J1535$-$571 \citep[][]{zhangEvolutionCoronaMAXI2022}. This model is able to explain the radiative properties of QPOs and can offer unique insights on the physical properties of the corona, by simultaneously fitting the time-averaged energy spectrum of the BH LMXB, and the energy-dependent rms amplitude and lags of the low-frequency QPOs.

In this paper, we study the fractional rms amplitude and phase-lag spectra of the type-B QPOs of the BH transient GX 339$-$4 during its 2021 outburst, observed for the first time over the 0.3$-$30 keV energy range with combined observations from two different satellites. Using the time-dependent Comptonization model developed by \citet{bellavitaVKompthVariableComptonization2022}, we fit the energy-dependent rms amplitude and lags of the type-B QPOs together with the energy spectrum of GX 339$-$4. Based on the results of this fit, we analyse the physical properties of the system and the geometry of the region or regions where the variability originates. In \cref{sec:obs}, we describe the observations and the analysis methods used to obtain the QPO properties and time-averaged spectral properties. In \cref{sec:spec-qpo-results}, we show the results of the spectral and timing analyses. In \cref{sec:model}, we describe the model we fit to the data. In \cref{sec:results-model}, we show the results of the model fitting and in \cref{sec:discussion}, we discuss the physical implications of these results.
 
\section{Observations and Data Analysis}
\label{sec:obs}
\subsection{Observations}
In January 2021, GX 339$-$4 went into outburst \citep[][]{tremouMeerKATObservationsRevealed2021} and was observed undergoing a hard-to-soft transition in late March of the same year \citep{liuInsightHXMTDetectionHardtosoft2021}. Here, we study \textit{NICER} \citep{gendreauNeutronStarInterior2016}, and \textit{AstroSat} \citep{singhASTROSATMission2014} observations of GX 339$-$4  during the SIMS of this outburst, when type-B QPOs appeared in the PDS of the source. The Large Area X-ray Proportional Counter \citep[\textit{LAXPC},][]{yadavLargeAreaXray2016} onboard \textit{AstroSat} has an energy range coverage of 3 to 80 keV, while the X-ray Timing Instrument (XTI) of \textit{NICER} has an energy coverage of 0.1 to 12 keV. Combining the data from both instruments, allowed us to cover a relatively broad energy band in the lag and fractional rms spectra. The \textit{AstroSat} and \textit{NICER} ObsIDs we analysed in this paper are listed in \cref{tab:obsids}.

To track the evolution of GX 339$-$4 during its 2021 outburst, we extracted the \textit{NICER} light curve of the source in the 0.75$-$12.0 keV band, between $\sim$59234 and $\sim$59573 MJD, using the \textsc{NICERDAS} pipeline and the \texttt{XSELECT} tool. We also extracted the light curves in two energy bands, 2.0$-$4.0 keV and 4.0$-$12.0 keV, to obtain the hardness ratio of the source and describe its spectral state during the outburst using a Hardness Intensity Diagram \citep[HID;][]{homanCorrelatedXraySpectral2001}.
\begin{table}
\caption{ObsIDs and time intervals with type-B QPOs of the \textit{AstroSat} and \textit{NICER} observations of GX 339$-$4 during the SIMS of its 2021 outburst.}
\label{tab:obsids}
\begin{tabular}{ccc}
    \hline
    ObsID & Start time (MJD) & Stop time (MJD)\\
    \hline
    \multicolumn{3}{c}{\textit{AstroSat}}\\
    \hline
    T03\_291T01\_9000004278 & 59303.60566354 & 59303.63589120\\
    & 59303.67332180 & 59303.70924306\\
    & 59303.80863256 & 59303.81524225\\
    & 59303.87772327 & 59303.89403935\\
    & 59303.91959491 & 59303.92206050\\
    & 59303.95047432 & 59303.95848380\\
    & 59303.98403935 & 59303.98970306\\
    \hline
    \multicolumn{3}{c}{\textit{NICER}}\\
    \hline
    4133010107 & 59303.59672454 & 59303.61059028\\
     & 59303.66127315 & 59303.67513889\\
     & 59303.79038194 & 59303.80427083\\
     & 59303.85495370 & 59303.85681713\\
     & 59303.85684028 & 59303.86541667\\
     & 59303.86543981 & 59303.86873843\\
     & 59303.91947917 & 59303.92608796\\
	 & 59303.92611111 & 59303.93332176\\
	 & 59303.98403935 & 59303.99789352\\
    \hline
\end{tabular}
\end{table}

\subsection{Spectral analysis}
\label{sub-sec:spectrum}
Using the \textsc{NICERDAS} pipeline, we extracted the \textit{NICER/XTI} energy spectrum of the source, during the time-intervals where the type-B QPO is present in the observations, to create one single energy spectrum (see \cref{sub-sec:QPOs}, for the method we used to define these time-intervals, and \cref{tab:obsids} for the time-intervals in MJD). The background was computed using the \texttt{nibackgen3C50} tool. In the analysis presented here, we excluded the \textit{XTI} energy spectrum below 1 keV due to the presence of edge features in the effective area of the instrument at low energies, e.g. the absorption edge feature of oxygen at 0.5 keV. These edge features especially affect the calibration of the energy spectrum of bright sources, like GX 339$-$4, producing abnormally large residuals in the spectral fit.

To extract the \textit{AstroSat/LAXPC} energy spectra we processed the observation T03\_291T01\_9000004278 from orbits 29755$-$29865 following the procedures described by the \textit{AstroSat} Science Support cell\footnote{\url{http://astrosat-ssc.iucaa.in/laxpcData}}, using the \texttt{Format (A)} version of the \textit{LAXPC} software. Specifically, we first converted the files from level 1 to level 2 using the commands \texttt{laxpc\_make\_filelist} and \texttt{laxpc\_make\_event}, and computed the good time intervals (GTI) during which the source was not occulted by the Earth and the satellite was outside the South Atlantic Anomaly using the command \texttt{laxpc\_make\_stdgti}. Subsequently we extracted the source and background spectra and created the response files during the time the type-B QPO was detected using the commands \texttt{laxpc\_make\_spectra} and \texttt{laxpc\_make\_backspectra}. Finally, we rebinned the source spectrum to oversample the intrinsic energy resolution of the detector by a factor of 3 and added a 2\% systematic error to the data. We only used data of the \textit{LAXPC} detector 2, which is the best calibrated detector of the \textit{LAXPC}.

Since the \textit{LAXPC} background spectrum is computed from non-simultaneous blank-sky observations, to check the validity of the data we plotted the total and background spectra on top of each other over the energy range $20-100$ keV. We expect that, since the detector is not sensitive to photons from the source above $\sim$50 keV, the total and background spectra should overlap above that energy. We found that in this case the total spectrum is always significantly brighter than the background spectrum up to the highest energy. Moreover, the difference between the two spectra is energy dependent, with the ratio of the two spectra being between $\sim$1.1 at 50 keV and $\sim$1.03 at 80 keV, which means that the background spectrum cannot be corrected by a simple multiplicative factor to match the total spectrum at high energies. Since the {\em LAXPC} data at energies above $\sim$12 keV weights heavily upon the best-fitting power-law index and electron temperature of the Comptonizing component (see \cref{sec:model} for details about the models we use to fit the energy spectrum of the source), and at this point the {\em LAXPC} spectra proved to be unreliable, we decided not to include the {\em LAXPC} spectrum in the rest of the analysis.

\subsection{Fourier timing analysis}
\label{sub-sec:QPOs}
For both the \textit{AstroSat} and \textit{NICER} observations of GX 339$-$4, we computed a Leahy-normalised \citep{leahySearchesPulsedEmission1983} dynamical PDS, using the  \texttt{GHATS}\footnote{\url{http://www.brera.inaf.it/utenti/belloni/GHATS_Package/Home.html}} analysis package, and selected the time-intervals when type-B QPOs are simultaneously observed by both instruments and appeared strong and narrow ($Q \gtrsim 7$) in the PDS. All the time-intervals containing QPOs are within a narrow range of hardness ratio during the state transition of the source, as we expect for type-B QPOs, defining this state as the SIMS of the outburst. In \cref{tab:obsids} we list the time-intervals in MJD with type-B QPOs that we identified for both the \textit{AstroSat} and \textit{NICER} observations. The data segments of the dynamical PDS have a length of $\sim$13 s and the time resolution is 400 $\mu$s, hence the minimum frequency and the frequency resolution are 0.08 Hz and the Nyquist frequency is 1250 Hz.

To study the timing properties of the type-B QPOs of GX 339$-$4, we extracted a fractional rms-squared normalised \citep[][]{belloniVariabilityNoiseProperties1990} PDS averaged only over the time-intervals with type-B QPOs – that we identified from the dynamical PDS. As for the dynamical PDS, the data segments of the PDS have a length of $\sim$13 s and the time resolution is 400 $\mu$s. We used a multi-Lorentzian model \citep[][]{belloniUnifiedDescriptionTiming2002} with four different components, to characterise the variability in the PDS of GX 339$-$4 \citep[similar to][]{nowakAreThereThree2000}: a Lorentzian component centred at zero that describes the low-frerquency noise (LFN) in the PDS, a strong narrow component, that we identified as the fundamental/type-B QPO, and two other weaker components that we identified as the sub-harmonic and the second harmonic of the QPO. We handle the Poisson noise present in the PDS by fitting a constant to the high frequency end of the spectra (frequencies higher than $\sim$300-500 Hz), where noise processes dominate, and subtracting it from the data. In those cases in which the amplitude of the Lorentzian component in the PDS of one of the instruments was not significant enough ($<3\sigma$) we report the 95\% upper limit, which we calculated by fixing the FWHM of the Lorentzian function to the value we obtained for the analogous component in the fit of the PDS of the other instrument (this was the case for the second harmonic component in the \textit{NICER} PDS and the sub-harmonic component in the \textit{AstroSat} PDS).

We calculated the rms-normalised PDS of the time-intervals with type-B QPOs for a set of energy bands to obtain the fractional rms spectrum of the QPO. For the \textit{NICER} data we used the $0.3-1.0$ keV, $1.0-1.5$ keV, $1.5-2.0$ keV, $2.0-3.0$ keV, $3.0-4.0$ keV, $4.0-6.0$ keV, $6.0-8.0$ keV and $8.0-12.0$ keV energy bands, and for the AstroSat data we used the $3.0-4.0$ keV, $4.0-6.0$ keV, $6.0-8.0$ keV, $8.0-12.0$ keV and $12.0-30.0$ keV energy bands. Using these same energy bands as subject bands and considering the $4.0-6.0$ keV and $3.0-4.0$ keV bands as the reference bands for the \textit{NICER} and \textit{AstroSat} data, respectively, we calculated the cross-spectra of the time-intervals with type-B QPOs to obtain the lag spectrum of the QPO. To minimise the errors in the lags, we chose as reference bands the energy bands that had the best combination of high count rate, at energies near the peak of the effective area of the instrument, and high amplitude of the variability. The lags were calculated averaging the real and imaginary parts of the cross-spectra of each energy band over one FWHM around the central frequency of the QPO, obtained from the fit to the corresponding PDS.

\begin{figure*}
    \centering
    \includegraphics[width=\linewidth]{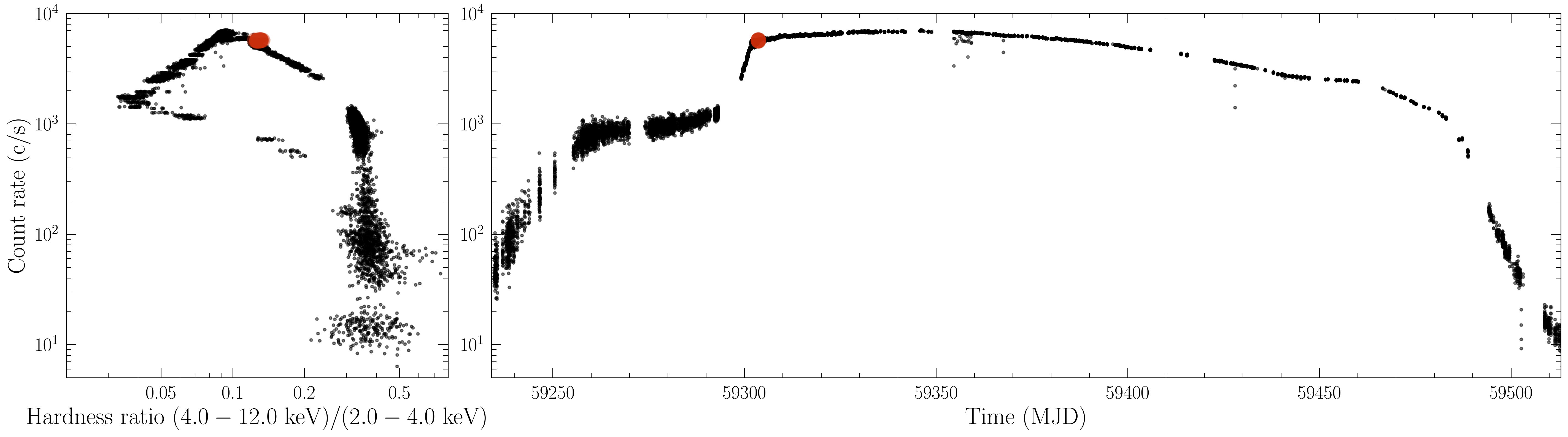}
    \caption{Hardness Intensity Diagram (HID; \textit{left panel}) and light curve (\textit{right panel}) of the 2021 outburst of GX 339$-$4 with \textit{NICER}. The intensity is defined as the count rate in the $0.75-12.0$ keV energy band and the hardness ratio is the ratio between the $4.0-12.0$ keV band and the $2.0-4.0$ keV band. Each point corresponds to $\sim$13 s time-intervals. The red circles indicate the time-intervals with type-B QPOs in the PDS.}
    \label{fig:lc-hid}
\end{figure*}
\begin{figure}
    \centering
    \includegraphics[width=\linewidth]{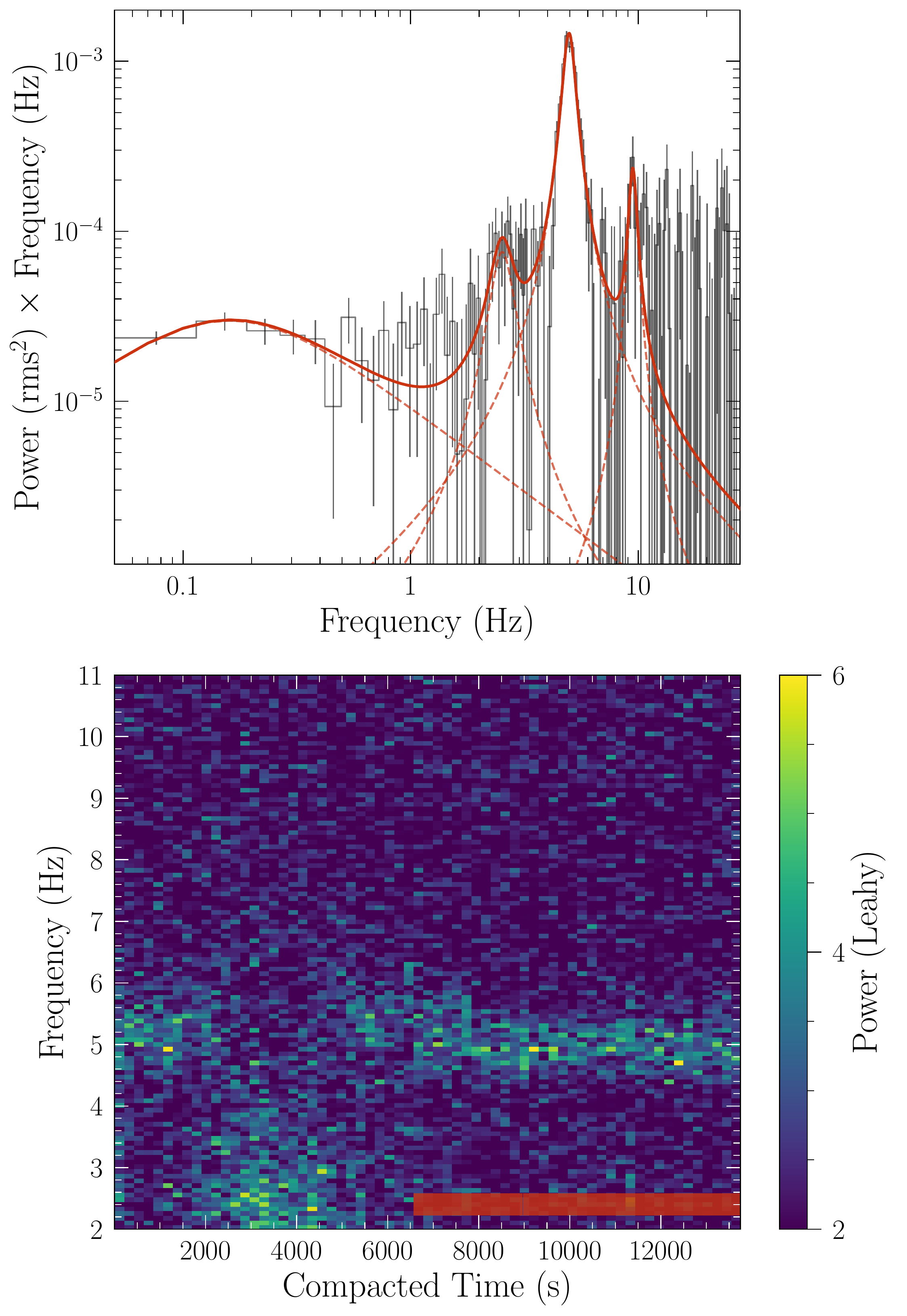}
    \caption{\textit{Top} panel: Averaged PDS of GX 339$-$4, with \textit{NICER} data in the 0.3$-$12 keV energy band, over the time-intervals when the type-B QPO is present in the observations. The solid red line indicates the best-fitting multi-Lorentzian model and the dashed red lines describe the individual components of the fit. \textit{Bottom} panel: Dynamical PDS of GX 339$-$4, with \textit{NICER} data in the 0.3$-$12 keV energy band rebinned to a time resolution of $\sim$210 s. The red band at the bottom of the plot indicates the time-intervals where the type-B QPO is present \textit{simultaneously} in the \textit{NICER} and \textit{AstroSat} observations (listed in \cref{tab:obsids}). All time gaps were removed from the data.}
    \label{fig:dyn-pds_nicer}
\end{figure}
\begin{figure}
    \centering
    \includegraphics[width=\linewidth]{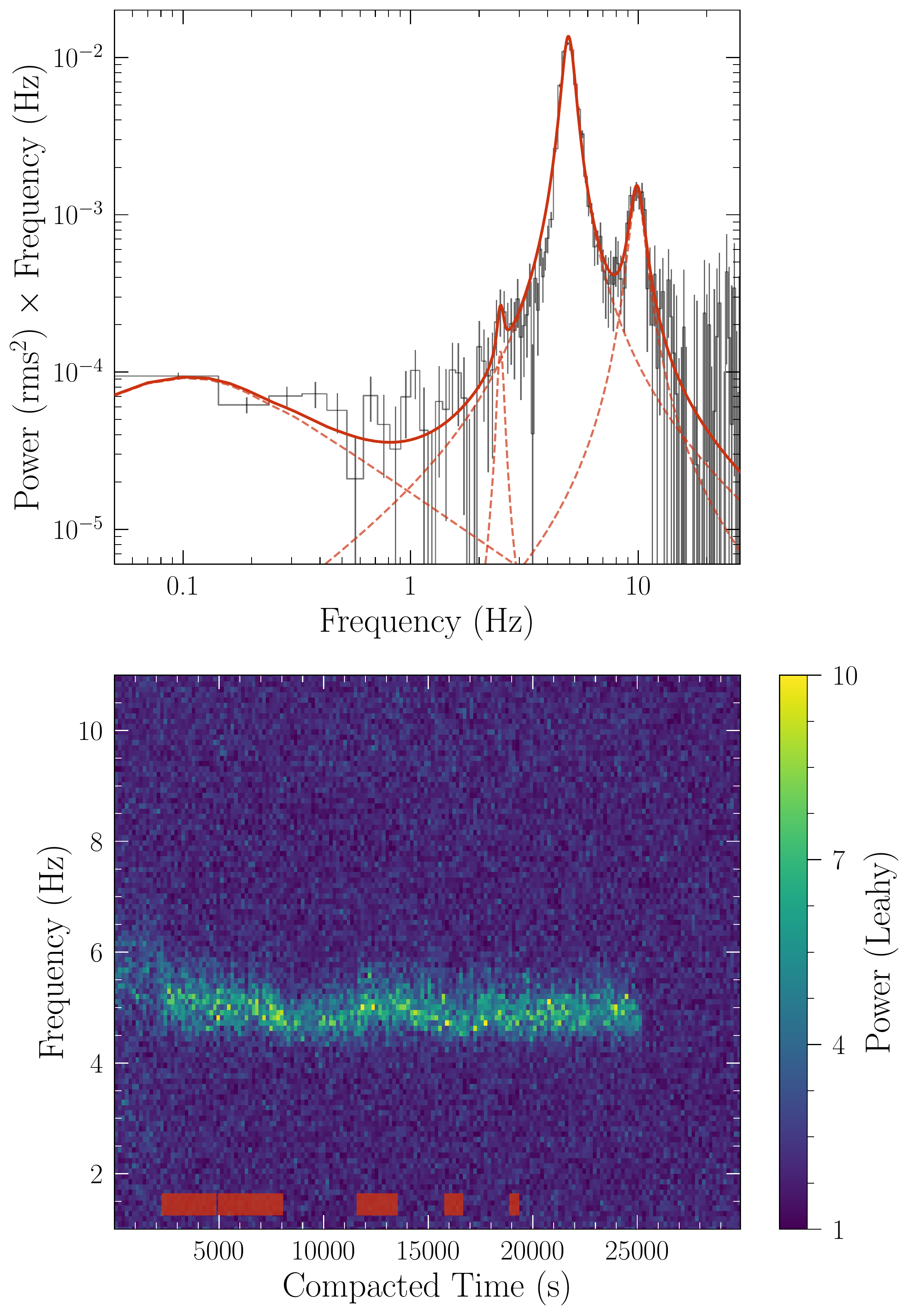}
    \caption{\textit{Top} panel: Same as \cref{fig:dyn-pds_nicer} for the \textit{AstroSat} data in the 3$-$60 keV energy band. \textit{Bottom} panel: Same as \cref{fig:dyn-pds_nicer} for the \textit{AstroSat} data in the 3$-$60 keV energy band rebinned to a time resolution of $\sim$160 s.}
    \label{fig:dyn-pds_astrosat}
\end{figure}
As mentioned in \cref{sub-sec:spectrum}, the background estimation of the \textit{AstroSat} data considerably influences the best-fitting parameters of the Comptonizing component in the energy spectrum of GX 339$-$4. Consequently, we did not take into account the \textit{LAXPC} energy spectrum in the rest of the analysis, but we did include the \textit{AstroSat} observations in the Fourier timing analysis of the type-B QPO, as the effect of the background is negligible in the computation of the rms-normalised PDS\footnote{For instance, in the case of this observation of GX 339$-$4 a change of only 5\% in the background of the source, that translates on a difference of a factor $\sim$3 in the relevant parameters of the Comptonizing component, produces variations in the fractional rms amplitude of at most a factor $\sim$1.05.} and has no weight in the calculation of the lags. Also, while we excluded the \textit{XTI} energy spectrum below 1 keV in the analysis of the source, we include the entire \textit{NICER} energy range in the calculation of both the lags and the fractional rms, as the calibration of the effective area of the instrument has no effect on these properties of the QPO.

\section{Spectral and Timing Products}
\label{sec:spec-qpo-results}
\subsection{Spectral results}
\label{sub-sec:spec-results}
Both the \textit{NICER} light curve and HID of GX 339$-$4 during the 2021 outburst are shown in \cref{fig:lc-hid}, where each data point corresponds to $\sim$13 s and the time-intervals with type-B QPOs in their PDS are indicated in red. The $0.75-12.0$ keV light curve in the right panel of \cref{fig:lc-hid} shows that, after the start of the outburst at $\sim$59234 MJD, the X-ray emission from the source rose from $\sim$10 c/s up to $\sim$8000 c/s at around $\sim$59350 MJD and then gradually decayed towards the end of the outburst at $\sim$59573. The observations with type-B QPOs appear shortly before the peak in the light curve, at around $\sim$59303 MJD.

During the outburst, GX 339$-$4 traces a \textit{"q-shape"} curve in the HID (left panel in \cref{fig:lc-hid}) as the $0.75-12.0$ keV count rate of the source increases and then decreases with time. While the source is in the hard state, at the beginning of the outburst, the hardness ratio remains more or less constant at $\sim$0.4 and the count rate increases from $\sim$10 c/s to $\sim$1000 c/s. During the transition from the hard to the soft state, while the source reaches the maximum intensity, the hardness ratio decreases from $\sim$0.3 to $\sim$0.05. At the end of the outburst, the count rate steadily decreases and the source returns to the hard state. In the HID, it is apparent that all the time-intervals with type-B QPOs appear in a narrow range in hardness ratio, which we identified as the SIMS of the outburst.

\subsection{Timing results}
\label{sub-sec:qpo-results}
\begin{table}
\caption{Best-fitting parameters and 1$\sigma$ errors of the fit to the \textit{NICER} and \textit{AstroSat} PDS of the the time-intervals with type-B QPOs.}
\label{tab:fit-PDS}
\begin{tabular}{lccc}
    \hline
    \multicolumn{4}{c}{\textit{NICER} PDS (0.3$-$12) keV}\\
    Component & $\nu_0$ (Hz) & FWHM (Hz) & rms (\%)\\
    \hline
    Fundamental & $4.95 \pm 0.01$ & $0.63 \pm 0.04$ & $1.70 \pm 0.03$\\
    Sub-harmonic & $2.5 \pm 0.1$ & $0.6 \pm 0.2$ & $0.53 \pm 0.08$\\
    Second harmonic & $9.7 \pm 0.3$ & 1.5$^\dag$ & $<0.73^{*}$\\
    LFN & 0$^\dag$ & $0.31 \pm 0.04$ & $0.96 \pm 0.03$\\
    \hline
    \multicolumn{4}{c}{\textit{AstroSat} PDS (3$-$60) keV}\\
    Component & $\nu_0$ (Hz) & FWHM (Hz) & rms (\%)\\
    \hline
    Fundamental & $4.93 \pm 0.01$ & $0.64 \pm 0.01$ & $5.29 \pm 0.03$\\
    Sub-harmonic & $2.4 \pm 0.4$ & 0.6$^\dag$ & $<0.62^{*}$\\
    Second harmonic & $9.48 \pm 0.07$ & $1.5 \pm 0.2$ & $1.85 \pm 0.08$\\
    LFN & 0$^\dag$ & $0.19 \pm 0.02$ & $1.66 \pm 0.04$\\
    \hline
    \multicolumn{4}{l}{$^\dag$ fixed parameters.}\\
    \multicolumn{4}{l}{$^{*}$ 95\% upper limits.}
\end{tabular}
\end{table}
In the \textit{bottom} panel of \cref{fig:dyn-pds_nicer}, we show the \textit{NICER} dynamical PDS of GX 339$-$4 in the 0.3$-$12 keV energy band, where the red band at the bottom of the plot represents the time-intervals when type-B QPOs are present simultaneously in the \textit{NICER} and \textit{AstroSat} observations (listed in \cref{tab:obsids}). Using exclusively these time-intervals we calculated the averaged \textit{NICER} PDS of the type-B QPO, shown in the \textit{top} panel of \cref{fig:dyn-pds_nicer}. In this figure, the best-fitting multi-Lorentzian model is shown in red with four different variability components, shown with red dashed lines, clearly identifiable: the LFN component, the fundamental/type-B QPO, the sub-harmonic and the second harmonic. Analogous to \cref{fig:dyn-pds_nicer}, in \cref{fig:dyn-pds_astrosat} we plot the \textit{AstroSat} PDS and dynamical PDS in the 3$-$60 keV energy band. In the \textit{top} panel of \cref{fig:dyn-pds_astrosat} we again show the best-fitting multi-Lorentzian model, that consists of four variability components, and in the \textit{bottom} panel of \cref{fig:dyn-pds_astrosat} the red bands represent the time-intervals with type-B QPOs simultaneously in the \textit{NICER} and \textit{AstroSat} observations. The best-fitting parameters to the \textit{NICER} and \textit{AstroSat} PDS and their 1$\sigma$ errors are listed in \cref{tab:fit-PDS}.

The fractional rms amplitude and phase-lag spectra of the type-B QPO of GX 339$-$4 are shown in the left and centre panels of \cref{fig:fit-dual}. In this figure it is apparent that the fractional rms amplitude remains more or less constant from 0.7 keV to $\sim$1.8 keV and then increases with increasing energy, from $\sim$1 per cent at 1.5$-$2.0 keV to $\sim$17 per cent at 20$-$30 keV. The energy-dependent phase-lags are all positive, decreasing from $\sim$1.2 rad at 0.7 keV to 0 rad at $\sim$3.5 keV and then increasing again to $\sim$0.6 rad at 20$-$30 keV. The phase-lag spectrum outlines a \textit{"U-shaped"} curve centred at the energy of the reference band, showing that, at the QPO frequency, the photons in all energy bands lag behind the photons in the 4$-$6 keV band for the \textit{NICER} data and in the 3$-$4 keV band for the \textit{AstroSat} data. 

\section{Model}
\label{sec:model}
In this paper, we fit simultaneously the time-averaged energy spectrum of GX 339$-$4, and the phase-lag and fractional rms amplitude spectra of the type-B QPO using the time-dependent Comptonization model \textsc{vkompth} developed by \citet{bellavitaVKompthVariableComptonization2022} for low-frequency QPOs in BH LMXBs, based on the model by \citet{karpouzasComptonizingMediumNeutron2020}\footnote{In principle \textsc{vkompth} can be used to model the radiative properties of other variability components in the PDS of LMXBs, e.g. the BBN, as long as that component has a characteristic frequency}. The \textsc{vkompth} model conceives QPOs as small oscillations around the solution of the stationary Kompaneets equation \citep[][]{kompaneetsEstablishmentThermalEquilibrium1957}, which describes the steady-state energy spectrum of the source. This way of defining the variability in the X-ray emission of BHXBs allows us to link the behaviour of the energy-dependent lags and rms amplitude with the physical properties of the system represented by the parameters of the Kompaneets equation.

The \textsc{vkompthdk} model considers that the soft photon source, i.e. the geometrically thin and optically thick accretion disc around the BH \citep[][]{shakuraBlackHolesBinary1973}, feeds low energy photons to a spherically symmetric and homogeneous corona of size $L$ and temperature $kT_e$, where the photons, before escaping, experience one or multiple inverse Compton scatterings. A fraction of these – now high energy – photons that escape the corona impinge back onto the accretion disc. This feedback process is parameterised in the model by the variable $\eta \in [0,1]$, which represents the fraction of the disc flux that is due to feedback from the corona. This feedback fraction $\eta$ is related to the fraction of coronal photons that return to the disc, the intrinsic feedback fraction, $\eta_{\mathrm{int}}$. The model internally calculates $\eta_{\mathrm{int}}$ through the relation $\eta = \eta_{\mathrm{int}}/\eta_{\mathrm{int,\, max}}$, where $\eta_{\mathrm{int,\, max}}$ – the maximum value of $\eta_{\mathrm{int}}$ – depends upon the parameters of the disc and corona \citep[see Appendix A in][]{karpouzasComptonizingMediumNeutron2020}. In these conditions, the model treats the oscillations in the energy spectrum – the QPOs – as perturbations of the coronal temperature, $kT_e$, and, via feedback, of the soft photon source temperature, $kT_s$, produced by variability of the external heating source $\delta \dot{H}_{\mathrm{ext}}$. This external heating source provides the energy that balances the cooling of the corona due to the inverse Compton scattering process \citep[see][for a more detailed explanation of the model and the solving scheme]{bellavitaVKompthVariableComptonization2022}.

\begin{table*}
\caption{Best-fitting parameters and $1\sigma$ errors of the joint fit of the fractional rms and phase-lag spectra of the type-B QPO and the time-averaged energy spectrum of GX 339-4 to the \textit{NICER} and \textit{AstroSat} data using a single-corona model.}
\label{tab:fit-single}
\begin{tabular}{lcccc@{\hspace*{20pt}}c@{\hspace*{20pt}}}
    \hline
    Component & \texttt{TBfeo} & \multicolumn{4}{c}{\texttt{nthComp}}\\
    \\
    Parameter & $N_\mathrm{H}$ ($10^{22}$cm$^{-2}$) & Flux (10$^{-8}$ erg/cm$^2$/s) & $\Gamma^*$ & $\tau$ & $kT_e$ (keV)$^*$\\
    \hline
    & $0.46 \pm 0.01$ & $1.9 \pm 0.1$ &  $3.68 \pm 0.03$ & $0.81$ & $33_{-10}^{+12}$\\
    \hline
    Component & \multicolumn{2}{c}{\texttt{diskbb}} & \multicolumn{3}{c}{\texttt{gaussian}}\\
    \\
    Parameter & Flux (10$^{-8}$ erg/cm$^2$/s) & $kT_\mathrm{in}$ (keV)$^*$ & Flux (10$^{-2}$ ph/cm$^2$/s) & E$_0$ (keV) & $\sigma$ (keV)\\
    \hline
    & $0.63 \pm 0.04$ & $0.66 \pm 0.01$ & $0.30 \pm 0.04$ & $6.54 \pm 0.05$ & $0.43 \pm 0.05$\\
    \hline
    Component & \multicolumn{5}{c}{\textsc{vkompthdk}}\\
    \\
    Parameter & $R_\mathrm{in}$ (km) & $L$ ($10^2$ km) & $\eta$ & $\eta_\mathrm{int}$ & $\delta \dot{H}_{\mathrm{ext}}$\\
    \hline
    & $250^\dag$ & $120 \pm 12$ & $0.46 \pm 0.04$ & $0.17 \pm 0.01$ & $0.22 \pm 0.01$\\
    \hline
    \multicolumn{6}{l}{$^*$ parameters linked to their corresponding parameter in \textsc{vkompthdk}. See text for details.}\\
    \multicolumn{6}{l}{$^\dag$ fixed parameters.}
\end{tabular}
\end{table*}
While fitting the phase-lag and rms spectra of the QPO and the energy spectrum of the source in \texttt{XSPEC} \citep[][]{arnaudXSPECFirstTen1996}, with \textsc{vkompthdk} compiled as an external model, we simultaneously characterised the time-averaged energy spectrum with a combination of models that describe the shape of the different components that dominate the emission of the source: \texttt{TBfeo*(diskbb+gaussian+nthComp)}. The model that we fit to the time-averaged energy spectrum of the source is parameterised by the hydrogen column density, $N_\mathrm{H}$, of the \texttt{TBfeo} component; the temperature at the inner disc radius, $kT_\mathrm{in}$, and the normalisation of the \texttt{diskbb} component; the line energy, E$_0$, the line width, $\sigma$, and the normalisation of the Fe-line \texttt{gaussian} component; and the power-law photon index, $\Gamma$, the electron temperature, $kT_e$, the seed-photon temperature, $kT_\mathrm{bb}$, and the normalisation of the \texttt{nthComp} component. Simultaneously, the \textsc{vkompthdk} model is characterised by the soft-photon source temperature, $kT_s$, the electron temperature, $kT_e$, the power-law photon index, $\Gamma$, the size of the corona, $L$, the feedback fraction, $\eta$, the inner disc radius, $R_\mathrm{in}$, and the variability of the external heating rate, $\delta \dot{H}_{\mathrm{ext}}$. During the fit, the seed-photon temperatures of \texttt{nthComp} and \textsc{vkompthdk}, $kT_\mathrm{bb}$ and $kT_s$, respectively, are linked to the temperature of \texttt{diskbb}, $kT_\mathrm{in}$. The electron temperature, $kT_e$, and power-law photon index, $\Gamma$, of \textsc{vkompthdk} are also linked during the fit to the corresponding parameters in \texttt{nthComp}. It is important to note that in the fit of the energy spectrum we could replace \texttt{nthComp} with \textsc{vkompthdk} in the model and obtain the same results, as both components are equivalent when dealing with time-averaged/non-variable data.

As in \citet{bellavitaVKompthVariableComptonization2022}, we added a \textit{dilution} correction factor to the joint model described in the previous paragraph. This \textit{dilution} takes into account the effect that the non-variable emission – from either the disc blackbody or the Fe-line Gaussian – has on the fractional rms amplitude of the QPO calculated in the model, assuming that the variability comes solely from the Comptonizing component, \texttt{nthComp}. The impact of the energy-dependent \textit{dilution} factor, in terms of the components of the model, \texttt{dilution = nthComp/(nthComp+diskbb+gaussian)}, is mostly at low energies, where the non-variable/soft components dominate.

Following \citet{garciaTwocomponentComptonizationModel2021} and \citet{bellavitaVKompthVariableComptonization2022}, we also investigate the case where the oscillations in the spectrum of the source take place in two Comptonization regions, instead of only one. \citet{bellavitaVKompthVariableComptonization2022} called this model \textsc{vkdualdk} and labelled the two regions as \textit{small} and \textit{large} corona, denoted with sub-indices 1 and 2, respectively. The dual-corona model is characterised by two sets of parameters that are equivalent to the ones in the single-corona model: the soft-photon source temperatures, $kT_{s,1}$ and $kT_{s,2}$, the electron temperatures, $kT_{e,\,1}$ and $kT_{e,\,2}$, the power-law indices, $\Gamma_1$ and $\Gamma_2$, the sizes of the coronae, $L_1$ and $L_2$, the feedback fractions, $\eta_1$ and $\eta_2$, the inner disc radius, $R_\mathrm{in}$, and the variability of the external heating rates, $\delta \dot{H}_{\mathrm{ext},1}$ and $\delta \dot{H}_{\mathrm{ext},2}$. An extra phase parameter, $\phi$, accounts for the relative phase of the oscillation of the two coronae in the model. During the fit of the dual-corona model, we consider the values of the power-law photon indices, $\Gamma_1$ and $\Gamma_2$, and the electron temperatures, $kT_{e,\,1}$ and $kT_{e,\,2}$, to be the same in both coronae and are linked to the corresponding parameters in \texttt{nthComp}. While $kT_{s,1}$ is linked to both $kT_\mathrm{in}$ of \texttt{diskbb} and $kT_\mathrm{bb}$ of \texttt{nthComp}, we allow $kT_{s,2}$ to vary freely to account for a possible difference of the temperature of the seed-photon source that illuminates the two coronae. Whenever the resulting best-fitting parameters of the models are not significant enough ($<3\sigma$), we report the 95\% upper limits.

\section{Model Fitting}
\label{sec:results-model}
In \cref{tab:fit-single} we list the best-fitting parameters and their $1\sigma$ errors of the joint fit to the \textit{NICER} and \textit{AstroSat} phase-lag and fractional rms amplitude spectra of the type-B QPO and the \textit{NICER} time-averaged energy spectrum of the source, using the \textsc{vkompthdk} model. In the table the parameters linked during the fit appear only once. The total $\chi^2_\nu$ of the fit and the $\chi^2$ of each separate data-set in the fit are shown in \cref{tab:chi-2}. We calculated the errors of the parameters running Markov Chain Monte Carlo (MCMC) simulations with 120 walkers and a length of $10^5$ steps. In \cref{apen:single-fit} we show a plot with the best-fitting model to the data and a corner plot of the MCMC simulations for a sample of parameters from the fit.

The physical picture that the best-fitting single-corona model provides for GX 339$-$4 is of a corona of $\sim$12000 km or $\sim$1000 $R_g$ with $R_g = GM_{\mathrm{BH}}/c^2$, considering a BH with a mass of $8 M_\odot$ \citep[][]{heidaMassFunctionGX2017} and electron temperature of $\sim$30 keV, where the oscillations of the energy spectrum originate. Considering that
\begin{equation}
\label{eq:tau}
    \tau = \sqrt{\frac{9}{4} + \frac{3}{[kT_e/m_ec^2][\left(\Gamma + 1/2\right)^2 - 9/4]}} - \frac{3}{2} \; \mathrm{,}
\end{equation}
where $m_e$ and $c$ are the rest mass of the electron and the speed of light, respectively, and with $\Gamma \approx 3.7$, the coronal temperature yields an optical depth of $\tau \approx 0.8$. Considering the optical depth we obtained from the fit we can estimate the Compton \textit{y-parameter} $y = (4kT_e/m_ec^2)\max(\tau,\,\tau^2)$, a dimensionless parameter that characterises Comptonization \citep[see][]{zeldovichXRayEmissionAccompanying1969,shapiroTwotemperatureAccretionDisk1976}. For the single-corona fit $y \approx 0.21$, which is consistent with the system being in an unsaturated Comptonization regime, where the variable Comptonization \textsc{vkompthdk} model we have used is appropriate. The single-corona fit with \textsc{vkompthdk} yields $\eta_\mathrm{int} \approx$ 0.17, such that $\sim$17\% of the coronal photons impinge back onto the accretion disc.
\begin{figure*}
    \centering
    \includegraphics[width=\linewidth]{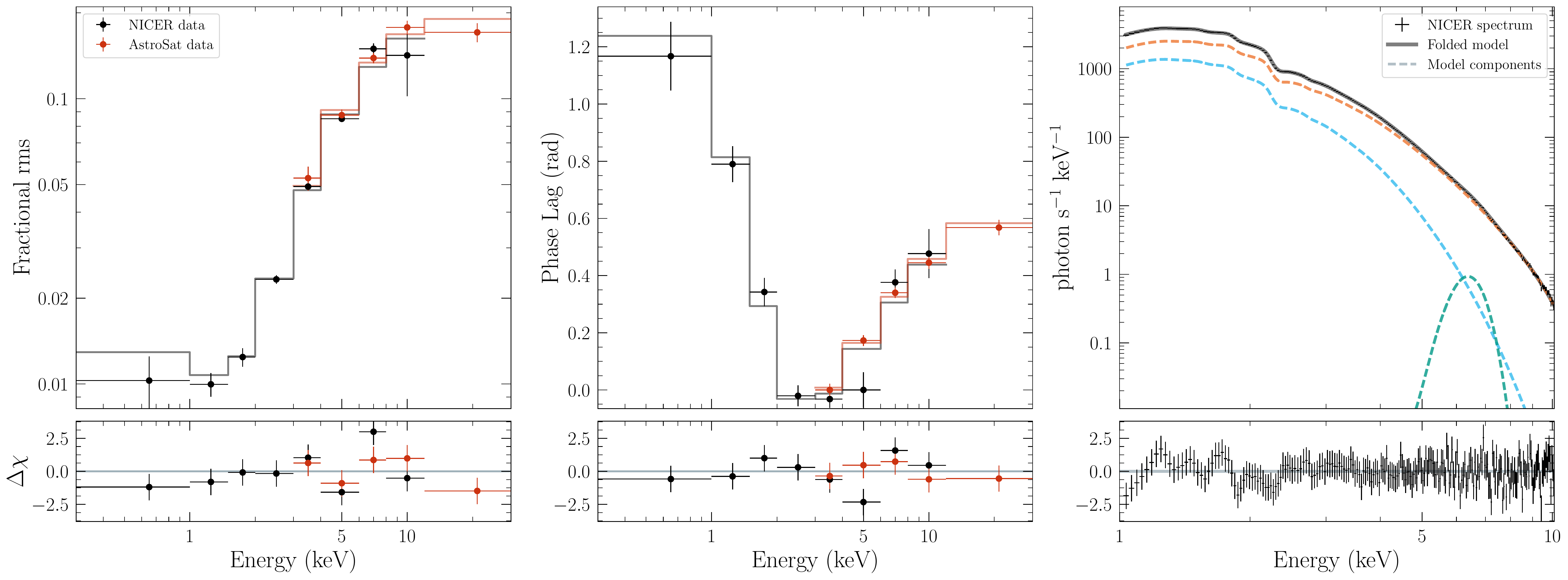}
    \caption{Fractional rms amplitude (\textit{left panel}) and phase-lag (\textit{centre panel}) spectra of the type-B QPO, and time-averaged energy spectrum (\textit{right panel}) of GX 339$-$4. Red and black correspond to AstroSat and \textit{NICER} data, respectively. The solid lines indicate the best-fitting model with a dual corona, \textsc{vkdualdk}, to the data. In the \textit{right panel}, the solid grey line corresponds to the total folded model used to fit the energy spectrum and the dashed lines correspond to the components of this model: disc blackbody (\textit{cyan}), \texttt{nthcomp} (\textit{orange}) and Gaussian Fe-line (\textit{teal}).}
    \label{fig:fit-dual}
\end{figure*}
\begin{table*}
\caption{Best-fitting parameters and $1\sigma$ errors of the joint fit of the fractional rms and phase-lag spectra of the type-B QPO and the time-averaged energy spectrum of GX 339-4 to the \textit{NICER} and \textit{AstroSat} data using a dual-corona model.}
\label{tab:fit-dual}
\begin{tabular}{lc@{\,}c@{\,}cc@{\hspace*{20pt}}c@{\hspace*{20pt}}c@{\hspace*{20pt}}}
    \hline
    Component & \texttt{TBfeo} & \multicolumn{4}{c}{\texttt{nthComp}} & \\
    \\
    Parameter & $N_\mathrm{H}$ ($10^{22}$cm$^{-2}$) & Flux (10$^{-8}$ erg/cm$^2$/s) & $\Gamma^*$ & $\tau$ & $kT_e$ (keV)$^*$ & \\
    \hline
    & $0.44 \pm 0.01$ & $1.78 \pm 0.04$ & $3.63 \pm 0.03$ & $0.4$ & $76 \pm 27$ &\\
    \hline
    Component & \multicolumn{2}{c}{\texttt{diskbb}} & \multicolumn{3}{c}{\texttt{gaussian}}\\
    \\
    Parameter & Flux (10$^{-8}$ erg/cm$^2$/s) & $kT_\mathrm{in}$ (keV)$^*$ & Flux (10$^{-2}$ ph/cm$^2$/s) & E$_0$ (keV) & $\sigma$ (keV) & \\
    \hline
    & $0.69 \pm 0.06$ & $0.68 \pm 0.01$ & $0.38 \pm 0.05$ & $6.48 \pm 0.05$ & $0.49 \pm 0.05$ &\\
    \hline
    Component & \multicolumn{6}{c}{\textsc{vkdualdk}} \\
    \\
    Parameter & $\phi$ (rad) & $kT_{s,\,2}$ (keV) & $R_\mathrm{in}$ (km) & $L_1$ ($10^2$ km) & $L_2$ ($10^2$ km) &\\
    \hline
    & $2.5 \pm 0.1$ & $0.41 \pm 0.01$ & $250^\dag$ & $2.9 \pm 0.6$ & $180 \pm 21$ & \\
    \\
    Parameter & $\eta_1$ & $\eta_2$ & $\eta_{\mathrm{int},\,1}$ & $\eta_{\mathrm{int},\,2}$ & $\delta \dot{H}_{\mathrm{ext},\,1}$ & $\delta \dot{H}_{\mathrm{ext},\,2}$\\
    \hline
    & $0.90 \pm 0.02$ & $<0.17^{**}$ & $0.33 \pm 0.01$ & $<0.04^{**}$ & $0.7 \pm 0.1$ & $<0.3^{**}$\\
    \hline
    \multicolumn{7}{l}{$^*$ parameters linked to their corresponding parameter in \textsc{vkdualdk}. See text for details.}\\
    \multicolumn{7}{l}{$^{**}$ 95\% upper limits.}\\
    \multicolumn{7}{l}{$^\dag$ fixed parameters.}
\end{tabular}
\end{table*}
\begin{figure*}
    \centering
    \includegraphics[width=\linewidth]{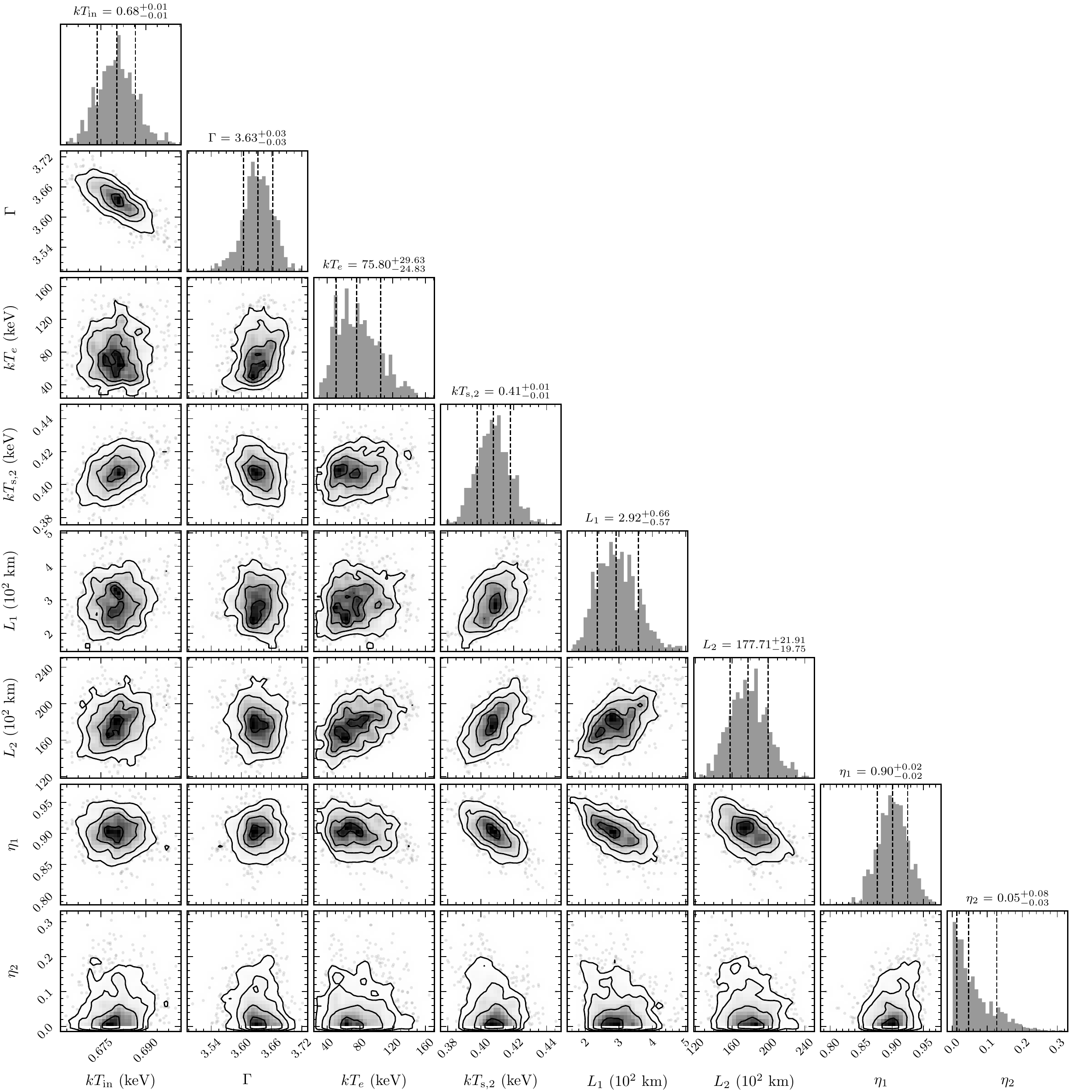}
    \caption{Corner plot of a selection of best-fitting parameters of the dual-corona fit to the fractional rms amplitude and phase-lag spectra of the type-B QPO, and the time-averaged energy spectrum of GX 339$-$4. The parameters, shown with their 1$\sigma$ errors in the title of each histrogram, are the temperature at the inner disc radius, $kT_\mathrm{in}$ (linked to $kT_\mathrm{bb}$ in \texttt{nthcomp} and $kT_{s,1}$ in \textsc{vkdualdk}), the power-law photon index, $\Gamma$ (linked to $\Gamma_1$ and $\Gamma_2$ in \textsc{vkdualdk}), the electron temperature, $kT_e$ (linked to $kT_{e,\,1}$ and $kT_{e,\,2}$ in \textsc{vkdualdk}), the seed-photon source temperature of the \textit{large} corona, $kT_{s,2}$, the sizes, $L_1$ and $L_2$, and the feedback parameters, $\eta_1$ and $\eta_2$, of both coronae. All the best-fitting parameters of the dual-corona fit are listed in \cref{tab:fit-dual}. The contours in the 2D histograms show the 68\%, 90\% and 95\% confidence levels. The dashed black lines in the diagonal histograms show the median value of the parameter and its 1$\sigma$ confidence levels.}
    \label{fig:MCMC-dual}
\end{figure*}

Despite the fact that the single-corona fit describes the data reasonably well, with $\chi^2_\nu = 1.36$, it is evident in \cref{fig:fit-single} that the \textit{NICER} fractional rms and phase-lag spectra, particularly below 3 keV, are not well characterised by this model. Specifically, the $\chi^2$ of the single-corona fit for the 8 spectral bins of the \textit{NICER} fractional rms and phase-lag spectra are 37.7 and 63.2, respectively. Additionally, while the single-corona fit describes fairly well the \textit{AstroSat} fractional rms spectrum, with a $\chi^2$ of 2.31 for the 5 spectral bins of data, it poorly characterises the \textit{AstroSat} lag spectrum, with a $\chi^2$ of 55.8. These values indicate that the goodness of the fit is dominated mostly by how well described is the time-averaged energy spectrum by the combination of a disc blackbody, a Gaussian and a Comptonization component, and does not reflect accurately enough the behaviour of the fractional rms and phase-lag spectra of the QPO. Considering the outcome of the single-corona fit and following \citet{garciaTwocomponentComptonizationModel2021}, we explore the possibility that the type-B QPO in GX 339$-$4 originates in two Comptonizing regions, using the dual-corona model \textsc{vkdualdk}.

The resulting best-fit of the dual-corona model is shown in \cref{fig:fit-dual}. The best-fitting parameters and their $1\sigma$ errors are listed in \cref{tab:fit-dual}, where, as in \cref{tab:fit-single}, the linked parameters appear only once and the errors are estimated using MCMC simulations. In \cref{fig:MCMC-dual} we show a corner plot with the MCMC simulation for a sample of resulting parameters from the fit. The panels on the diagonal of the corner plot show the posterior probabilities of the parameters of the model, with dashed black lines indicating the median value and the $1\sigma$ confidence levels. The rest of the panels in \cref{fig:MCMC-dual} show 2D histograms of each pair of parameters. Again, the total $\chi^2_\nu$ of the fit and the $\chi^2$ of each separate data-set in the fit are shown in \cref{tab:chi-2}. A visual comparison between \cref{fig:fit-single} and \cref{fig:fit-dual} shows that the dual-corona model describes the lower energy end of the fractional rms and phase-lag spectra significantly better than the single-corona one. In quantitative terms, the total $\chi^2_\nu$ of the fit decreases from 1.36 for the single-corona model to 0.88 for the dual-corona model, while the $\chi^2$ of the dual-corona fit for the 8 spectral bins of the \textit{NICER} fractional rms and phase-lag spectra decreased to 14.9 and 10.05, respectively, and the $\chi^2$ for the 5 spectral bins of the \textit{AstroSat} phase-lag spectrum decreased to 1.55.

The new picture that the dual-corona depicts is one of a \textit{small} corona, of $\sim$300 km ($\sim$25 $R_g$), with an intrinsic feedback fraction of $\eta_{\mathrm{int},\,1}\approx0.33$, and a \textit{large} corona, of $\sim$18000 km ($\sim$1500 $R_g$), with an intrinsic feedback fraction of $\eta_{\mathrm{int},\,2}\lesssim0.04$. This result means that the fraction of high energy photons from the \textit{small} corona that return to the accretion disc is considerably larger than that of the \textit{large} corona. The temperatures of both coronae, which are linked during the fit, are $\sim76$ keV, that together with $\Gamma \approx 3.6$, yields an optical depth $\tau \approx 0.4$. Considering this optical depth, the Compton \textit{y-parameter} of the dual-corona fit is $y \approx 0.24$, consistent with the system being in an unsaturated Comptonization regime, comparable to the results from the single-corona fit.

For both the \textsc{vkompthdk} and the \textsc{vkdualdk} models, we fixed the inner disc radius to $R_\mathrm{in} = 250$ km. This value is consistent with that obtained from the normalisation of the \texttt{diskbb} component in the fit to the time-averaged energy spectrum, $\mathrm{norm}_\mathrm{dbb}$, $R_\mathrm{in} = f_\mathrm{color} \times \sqrt{(\mathrm{norm}_\mathrm{dbb}D_{10})/\cos{\theta}} \approx 200 - 300$ km, with $D_{10} \approx 0.8 - 1.2$, the distance to GX 339$-$4 in units of 10 kpc \citep[][]{hynesDistanceInterstellarSight2004}, $\theta \approx 40 - 60\degree$, the inclination of the source \citep[][]{furstComplexAccretionGeometry2015} and $f_\mathrm{color} \approx 2$, the color correction factor \citep[][]{shimuraSpectralHardeningFactor1995,davisRelativisticAccretionDisk2005}. Since both the fractional rms amplitude and the lags are ratios of quantities proportional to the flux of the source – which is in turn proportional to $R_\mathrm{in}^2$ – neither \textsc{vkompthdk} nor \textsc{vkdualdk} are sensitive to variations of $R_\mathrm{in}$. Allowing $R_\mathrm{in}$ to vary freely during the fit instead of fixing it to a value consistent with the parameters of the black-body component does not significantly improve the results of the fit\footnote{For instance, fixing $R_\mathrm{in}$ to either 200 or 300 km changes the values of the relevant best-fitting parameters by just 0.1\%.}. 

\begin{table}
\caption{$\chi^2$ statistics for the joint fit of the fractional rms and phase-lag spectra of the type-B QPO and the time-average energy spectrum of GX 339$-$4 to the \textit{NICER} and \textit{AstroSat} data using the single and dual-corona models.}
\label{tab:chi-2}
\begin{tabular}{lcccc}
    \hline
    & \multicolumn{2}{c}{Single-corona} & \multicolumn{2}{c}{Dual-corona}\\
    & \multicolumn{2}{c}{$\chi^2$ (spectral bins)} & \multicolumn{2}{c}{$\chi^2$ (spectral bins)}\\
    \hline
    & \textit{NICER} & \textit{AstroSat} & \textit{NICER} & \textit{AstroSat}\\
    \hline
    Energy Spectrum & 201.1 (253) & & 195.5 (253) & \\
    %Energy & \multirow{2}{*}{201.4 (253)} & & \multirow{2}{*}{195.6 (253)} &\\
    %Spectrum & \\
    Fractional rms & 37.7 (8) & 2.3 (5) & 14.9 (8) & 5.1 (5)\\
    Phase-lag & 63.2 (8) & 55.8 (5) & 10.1 (8) & 1.6 (5)\\
    \hline
    Total $\chi^2_\nu$ (d.o.f.) & \multicolumn{2}{c}{1.36 (264)} & \multicolumn{2}{c}{0.88 (259)}\\
    \hline
\end{tabular}
\end{table}
\section{Discussion}
\label{sec:discussion}
We studied the rms and phase-lag spectra of the type-B QPOs in the BH LMXB GX 339$-$4 during the 2021 outburst of the source. The energy-dependent fractional rms amplitude remains more or less constant at energies lower than $\sim$1.8 keV and then increases with increasing energy, from $\sim$1 per cent at 1.5$-$2.0 keV to $\sim$17 per cent at 20$-$30 keV. The phase-lag spectrum is \textit{"U-shaped"}, decreasing from $\sim$1.2 rad at 0.7 keV to 0 rad at $\sim$3.5 keV (about the reference band energy), and increasing again above that energy to $\sim$0.6 rad at 20$-$30 keV. We performed a joint fit of the time-averaged energy spectrum of the source and the fractional rms and phase-lag spectra of the QPO using the time-dependent Comptonization model \textsc{vkompthdk} \citep[][]{karpouzasComptonizingMediumNeutron2020,bellavitaVKompthVariableComptonization2022}. Using a single Comptonizing component, we find a fair agreement of the model with the data at high energies, however, the model shows large residuals at low energies. A fit with a two-coronae model \citep[][]{garciaTwocomponentComptonizationModel2021,bellavitaVKompthVariableComptonization2022}, provides a significantly better fit to the data and shows that two physically-conected Comptonizing regions can explain the fractional rms amplitude and the phase-lags of the type-B QPO of GX 339$-$4 simultaneously: a \textit{small} corona with a high feedback fraction of photons returning from the corona to the disc and a \textit{large} corona with almost no feedback. 

\subsection{The origin of the rms and lags}
Comptonization in a geometrically thick and optically thin region – the corona – located in the vicinity of the compact object is considered to be the source of the hard component observed in the energy spectrum of LMXBs \citep[][]{sunyaevHardXraySpectrum1979,sunyaevComptonizationXraysPlasma1980}. Given that the amplitude of the variability in the power spectra of LMXBs increases with increasing energy \citep[see e.g.][]{sobolewskaSpectralFourierAnalyses2006,mendezPhaseLagsHighfrequency2013}, Comptonization, that dominates at high energies, must therefore be responsible for the radiative properties of these variability components. Particularly for low-frequency QPOs in BH systems, \citet{sobolewskaSpectralFourierAnalyses2006} showed that while the energy spectrum of a sample of BH sources contains a soft component associated with the accretion disc, the rms spectrum does not need that component 
\citep[see also][and references therein, for a similar discussion about high-frequency QPOs in BH LMXBs]{mendezPhaseLagsHighfrequency2013}. We observe a similar behaviour of the energy-dependent fractional rms amplitude of the type-B QPO in GX 339$-$4. In the leftmost panel of \cref{fig:fit-dual} the fractional rms amplitude remains constant below $\sim$1.8 keV, but then increases steeply with increasing energy, reaching $\sim$17 per cent at 20 keV. At these high energies the corona dominates the emission of the source and in this context, regardless of whether the mechanism that produces the variability takes place somehow in the accretion disc, the signal still needs to be amplified in the corona.

The interaction between the soft photons emitted by the accretion disc and the corona, where these photons are inverse-Compton scattered, would naturally produce hard lags \citep[][]{miyamotoDelayedHardXrays1988}. The phase-lag spectrum of the type-B QPO of GX 339$-$4 describes instead a \textit{"U-shaped"} curve, with photons at energies below $\sim$4 keV – about the energy of the reference bands of both instruments – lagging behind those at $\sim$4 keV (see \textit{centre} panel in \cref{fig:fit-dual}). Note that the magnitude of the lags at energies below $\sim$4 keV is overall larger than the magnitude of the lags at energies above $\sim$4 keV. A similar behaviour was observed for the type-B QPO in MAXI J1348$-$630, where the phase-lags decrease from $\sim$0.9 rad at $\sim$0.9 keV to 0 rad at $\sim$2.2 keV and then increase at higher energies up to $\sim$0.6 rad at 9 keV \citep[][]{belloniTimeLagsTypeB2020,garciaTwocomponentComptonizationModel2021}. Using a flat seed-photon spectrum emitting exclusively between 2 and 3 keV, \citet{belloniTimeLagsTypeB2020} proposed that the soft lags at energies below $\sim$2 keV are due to Compton down-scattering in the corona of the photons emitted by the disc. However, this flat seed-photon spectrum fails to account for the dilution that directly emitted photons, originated in the disc at energies lower than 2 keV, would produce at the lower-energy end of the phase-lag spectrum. Indeed, if one considers a more realistic seed spectrum, e.g. a disc blackbody, the soft photons that escape without being scattered dilute the lags of the Compton down-scattered photons, leading to a flat lag spectrum below $\sim$4 keV \textcolor{black}{(Kylafis, priv. comm.)}, contrary to what is observed.

The soft-lags at energies below the 4 keV can be explained if we consider a feedback loop, where a fraction of the photons emitted by the corona return to the accretion disc and are re-emitted at later times and at lower energies than the hard photons directly coming from the corona \citep[][]{leeComptonizationQPOOrigins1998,leeComptonUpscatteringModel2001,kumarEnergyDependentTime2014}. The Comptonization model \textsc{vkompth} \citep[][]{karpouzasComptonizingMediumNeutron2020,bellavitaVKompthVariableComptonization2022} includes this feedback loop between the corona and the disc, conceiving the QPO as an oscillation in the physical properties of the corona and the Comptonised emission at the frequency of the QPO. Using this Comptonization model, \citet{garciaTwocomponentComptonizationModel2021} showed that the \textit{"U-shaped"} phase-lag spectrum of the type-B QPO in MAXI J1348$-$630 can be explained considering feedback between disc and corona, accounting for both the soft lags at low energies and the hard lags at high energies, which are explained by photons from the disc being inverse-Compton scattered in the corona. The phase-lag spectra of the type-B QPOs in MAXI J1820$+$070 \textcolor{black}{(Saina et al., in prep.; Ma et al., in prep.)} and MAXI J1535$-$571 \citep[][]{zhangEvolutionCoronaMAXI2022} behave similarly to what we observe for GX 339$-$4 and MAXI J1348$-$630, with the lag spectrum displaying a \textit{"U-shaped"} curve. This behaviour indicates that the origin of the type-B QPOs in all these sources is the same and is indeed related to the corona.

The \textsc{vkompth} models do not explain the origin of the oscillation of the emission in LMXBs, but assume that such oscillation exists specifically to compute its fractional rms and lag spectra. As far as the model concerns, the QPOs could be due to relativistic precession of the inner regions of the accretion disc \citep[][]{stellaLenseThirringPrecessionQuasiperiodic1998,stellaCorrelationsQuasiperiodicOscillation1999} or an inner percessing torus \citep[][]{ingramLowfrequencyQuasiperiodicOscillations2009,fragileHighfrequencyTypeCQPOs2016}, disc-trapped oscillations or corrugation modes \citep[][]{katoTrappedRadialOscillations1980,wagonerRelativisticDiskoseismology1999,katoBasicPropertiesThinDisk2001}, etc. Recently, \citet[][]{mastichiadisStudyNaturalFrequencies2022}, using a simple model to describe the dynamical coupling of the hard coronal photons with the soft radiation from the accretion disc, suggested that QPOs in BH LMXBs could be produced by a resonant coupling between a hot Comptonizing corona and the accretion disc, in agreement with what we argue here.

While the Comptonization model proposed by \citet{leeComptonUpscatteringModel2001} is able to explain the large amplitude of the energy-dependent fractional rms of the QPO at high energies, it also predicts that the fractional rms must reach a minimum at a certain "pivot" energy\footnote{The variable energy spectrum, that emerges from adding a sinusoidal oscillation to the linearised Kompaneets equation \citep[][]{leeComptonizationQPOOrigins1998}, pivots around this energy.} to then increase again at lower energies. This pivot energy is proportional to the temperature of the soft-photon source, and is generally below 2$-$3 keV \citep[see][]{bellavitaVKompthVariableComptonization2022}. When the Comptonization model considers the surface of the NS – a blackbody – as the seed-photon source, it predicts that the rms amplitude of the QPO must increase at both high and low energies, with a distinct minimum. At the time \citet{leeComptonUpscatteringModel2001} predicted this feature of the variability of LMXBs, data at such low energies were unavailable, making impossible to test the actual behaviour of the fractional rms spectrum. The \textsc{vkompthdk} model \citep[][]{bellavitaVKompthVariableComptonization2022} assumes instead that the seed-photon source in BH LMXBs is a disc black-body, whose emission at low energies is higher than that of a blackbody, effectively diluting the variability and producing a flat fractional rms spectrum at energies lower than the pivot energy. This is exactly what we observe in our data, as the fractional rms spectrum of the type-B QPO in GX 339$-$4 remains more or less constant at energies lower than $\sim$1.8 keV, before steeply increasing at higher energies. This phenomenon, predicted by the variable Comptonization model we applied to GX 339$-$4 in this work, has been observed in other LMXBs \citep[see e.g.][]{casellaStudyLowfrequencyQuasiperiodic2004,belloniTimeLagsTypeB2020, garciaTwocomponentComptonizationModel2021,zhangEvolutionCoronaMAXI2022}.

\subsection{The geometry of the Comptonizing region}
Our best-fitting model using a single-corona (\textsc{vkompthdk}; see \cref{fig:fit-single}), yields a corona of $\sim$11900 km and an intrinsic feedback fraction of $\eta_{\mathrm{int}} = 17$\%. Despite the fact that this single-corona model can somewhat reproduce the overall shape of the fractional rms and phase-lag spectra of the QPO, at lower energies the best-fitting parameters are not consistent with the data and the fit has a $\chi_\nu = 1.36$ for 264 d.o.f. This discrepancy between the model and the data could be due to the simplified assumptions that the model makes about the properties of the corona, i.e. a spherically symmetric corona, with constant temperature and optical depth. Assuming that the variability takes place in two different, but physically connected, Comptonizing regions instead, using \textsc{vkdualdk}, provides a significantly better fit to the data, with $\chi_\nu = 0.85$ for 259 dof.

The best-fitting parameters of the \textsc{vkdualdk} model are specially interesting, as the properties of the two resulting coronae are seemingly quite different. The \textit{small} corona, with a size of $\sim$300 km, has a very high intrinsic feedback fraction $\eta_{\mathrm{int},1} \approx 33$\%, in contrast to the \textit{large} corona, of $\sim$18000 km, that has a very low intrinsic feedback fraction $\eta_{\mathrm{int},2} \lesssim 4$\%. The seed-photon source temperature of the \textit{large} corona is $\sim$0.4 keV, lower than that of the \textit{small} corona, which is $\sim$0.7 keV. To explain these coronal parameters, we can conceive a scenario where Comptonization occurs in a compact region located very close to the BH – the \textit{small} corona – interacting with photons emitted by the innermost regions of the accretion disc, and in a vertically extended region – the \textit{large} corona – that, to a lesser degree, interacts with the outer/colder parts of the disc \citep[see also][]{garciaTwocomponentComptonizationModel2021,bellavitaVKompthVariableComptonization2022,zhangEvolutionCoronaMAXI2022,mendezCouplingAccretingCorona2022}. These results indicate that the \textit{large} corona dominates the short time-scale variability, driving the lag spectrum of the QPO, while the \textit{small} corona impacts on the steady-state spectrum of the source. The low feedback fraction of the \textit{large} corona can be explained by associating this region with the base of a relativistic jet or outflow, where Comptonization occurs \citep[see e.g.][]{fenderQuenchingRadioJet1999,markoffGoingFlowCan2005,reigJetModelGalactic2015,reigPhotonindextimelagCorrelationBlack2018,kylafisCorrelationTimeLag2018,reigInclinationEffectsXray2019,wangDiskCoronaJet2021,wangNICERReverberationMachine2022} and whose onset is believed to take place during the state transitions of BH transients like GX 339$-$4 \citep[][]{belloniEvolutionTimingProperties2005}.

As we explained in \cref{sec:model}, the \textsc{vkdualdk} model \citep[][see also \citealt{karpouzasComptonizingMediumNeutron2020}]{bellavitaVKompthVariableComptonization2022} solves the time-dependent Kompaneets equation \citep[][]{kompaneetsEstablishmentThermalEquilibrium1957} under a number of assumptions that, while necessary to solve the equations, may pose some limitations on the interpretation of the fitting results: (1) The corona is spherically symmetric and homogeneous of size $L$, with uniform density and electron temperature; (2) the soft-photon spectrum is that of either a blackbody or an optically thick, geometrical thin, accretion disc, and is emitted as in \citet[][see also \citealt{zdziarskiBroadbandGrayXray1996,zycki1989MayOutburst1999}]{sunyaevComptonizationXraysPlasma1980}. Furthermore, the model assumes that there is an external source of energy, $\dot{H}_{\mathrm{ext}}$, that balances the Compton cooling and keeps the corona in equilibrium. In the model the temperature of the seed-photon source, the electron temperature of the corona, and the rate at which energy is supplied to the corona, oscillate at the QPO frequency with amplitudes $\delta kT_s$, $\delta kT_e$ and $\delta \dot{H}_{\mathrm{ext}}$, respectively\footnote{In future versions, the model will include oscillations of the size or, equivalently, the density and optical depth of the corona, and will allow for other geometries.}. Considering that Comptonization can take place in two coronae instead of one, as we do here when using the \textsc{vkdualdk} model, provides a step forward to approximating the likely complex geometry of the accretion flow responsible for the variability in LMXBs.

Even though the \textsc{vkompth} models assume that Comptonization takes place in spherical coronae, considering one of these regions to be vertically extended would still give a physically appropriate estimate of the system geometry. Indeed, given that the cross-section of the inverse Compton scattering decreases at higher energies - i.e. small-angles (forward) scattering dominates -, most of the scatterings occurring in an optically thin spherical region around the BH will be along the vertical direction. In other words, the size of the \textit{large} corona we obtained from the fit of the \textsc{vkdualdk} model could effectively represent a cylindrical rather than a spherical region, which is consistent with the vertically extended corona we propose here.

As discussed above, the link between the jet and the mechanism responsible for type-B QPOs in BH LMXBs has been already proposed based on the coupling of the state transitions of these sources with the launch of relativistic ejecta \citep[e.g.][]{soleriTransientLowfrequencyQuasiperiodic2008,fenderJetsBlackHole2009,russellDiskJetCoupling20172019,homanRapidChangeXRay2020}, and on the dependence of the timing-properties of the variability with the inclination of the source \citep[see e.g.][]{mottaGeometricalConstraintsOrigin2015,heilInclinationdependentSpectralTiming2015, vandeneijndenInclinationDependenceQPO2017,reigInclinationEffectsXray2019}. Specifically for the type-B QPOs in GX 339$-$4, \citet{kylafisQuantitativeExplanationTypeB2020} proposed a model where Comptonization occurs in a precessing jet that produces the variability in the photon-number spectral index of the energy spectrum of the source observed by \citet{stevensPhaseresolvedSpectroscopyType2016}. However, these oscillations of the power-law photon index, $\Gamma$, in the cycle of the QPO are also expected in the context of the \textsc{vkompth} models, that conceive QPOs as perturbations of the coronal temperature, $kT_e$. Indeed, if we consider that in the \textsc{vkompthdk} model the optical depth of the corona, $\tau$, remains constant during the fit and that $kT_e$, $\Gamma$ and $\tau$ follow the relation in \cref{eq:tau}, oscillations of $kT_e$ would automatically produce variability in $\Gamma$ without the need for a precessing jet \citep[see e.g. Fig. 4 in][]{karpouzasComptonizingMediumNeutron2020}. To understand the actual contribution of the relativistic jet in originating the type-B QPOs in GX 339$-$4, it would be useful to also consider oscillations in the optical depth, or equivalently the size, of the corona in the \textsc{vkompth} models in the future.

Although the \textsc{vkompth} model is conceived as a simplified picture of the complex phenomena taking place in the innermost regions of LMXBs, it is remarkable that the time-dependent Comptonization model of \citet{karpouzasComptonizingMediumNeutron2020} and \citet{bellavitaVKompthVariableComptonization2022} can successfully fit, at the same time, the fractional rms and lag spectra of the QPOs, and the time-averaged energy spectrum of these sources from 0.3 to 100 keV \citep[e.g.][]{garciaTwocomponentComptonizationModel2021,garciaEvolvingPropertiesCorona2022,karpouzasVariableCoronaGRS2021,mendezCouplingAccretingCorona2022,zhangEvolutionCoronaMAXI2022}. More studies like the one presented in this paper, applying this same model to other LMXBs, may help us understand the role that Comptonization actually plays in originating or modulating the variability in accreting LMXBs, and elucidate the true nature of the geometry of these sources and of the mechanism responsible for the variability we observe.

%\section{Conclusions}

%The last numbered section should briefly summarise what has been done, and describe the final conclusions which the authors draw from their work.

\section*{Acknowledgements}

The authors wish to thank the referee for constructive comments that helped improve the manuscript. This work is part of the research programme Athena with project number 184.034.002, which is (partly) financed by the Dutch Research Council (NWO). FG is a CONICET researcher. FG acknowledges support by PIP 0113 (CONICET) PICT-2017-2865 (ANPCyT), and PIBAA 1275 (CONICET). TMB acknowledges financial contribution from PRIN-INAF 2019 N.15. This research has made use of data and software provided by the High Energy Astrophysics Science Archive Research Center (HEASARC), which is a service of the Astrophysics Science Division at NASA/GSFC. This publication uses the data from the AstroSat mission of the Indian Space Research Organisation (ISRO), archived at the Indian Space Science Data Centre (ISSDC).

%%%%%%%%%%%%%%%%%%%%%%%%%%%%%%%%%%%%%%%%%%%%%%%%%%
\section*{Data Availability}
The data underlying this article are publicly available at the website of the High Energy Astrophysics Science Archive Research Center (HEASARC, \url{https://heasarc.gsfc.nasa.gov/}) and at the Astrobrowse (\textit{AstroSat} archive) website (\url{https: //astrobrowse.issdc.gov.in/astro_archive/archive}) of the Indian Space Science Data Center (ISSDC). The \textsc{vkompth 1.1.1} model, used in this paper, is publicly available in a Github repository (\url{https://github.com/candebellavita/vkompth}). The corner plots shown in this paper were created using \texttt{pyXspecCorner} (\url{https://github.com/garciafederico/pyXspecCorner}).



\bibliographystyle{mnras}
\bibliography{main} % if your bibtex file is called example.bib

%\begin{thebibliography}{99}
%\end{thebibliography}

%%%%%%%%%%%%%%%%%%%%%%%%%%%%%%%%%%%%%%%%%%%%%%%%%%

%%%%%%%%%%%%%%%%% APPENDICES %%%%%%%%%%%%%%%%%%%%%

\appendix
\section{Single-corona fit}
\label{apen:single-fit}
\begin{figure*}
    \centering
    \includegraphics[width=\linewidth]{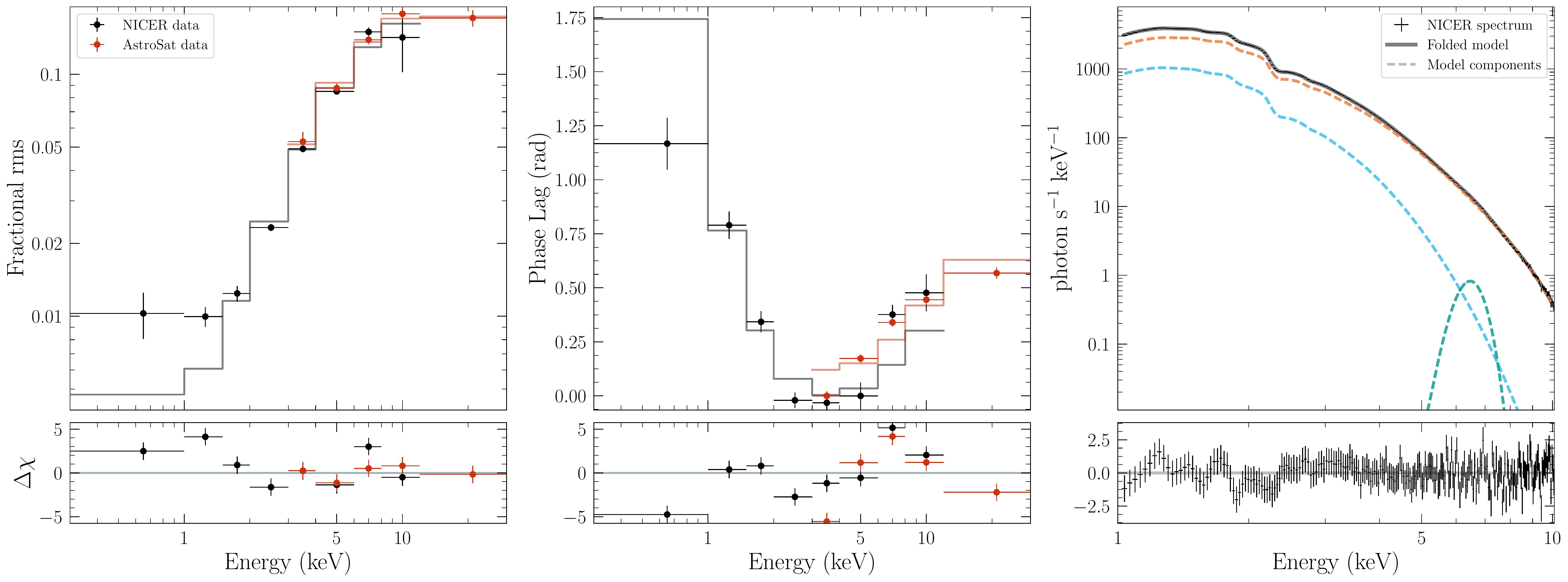}
    \caption{Fractional rms amplitude (\textit{left panel}) and phase-lag (\textit{centre panel}) spectra of the type-B QPO, and time-averaged energy spectrum (\textit{right panel}) of GX 339$-$4. Red and black correspond to AstroSat and \textit{NICER} data, respectively. The solid lines indicate the best-fitting model with a single corona, \textsc{vkompthdk}, to the data. In the \textit{right panel}, the solid grey line corresponds to the total folded model used to fit the energy spectrum and the dashed lines correspond to the components of this model: disc blackbody (\textit{cyan}), \texttt{nthcomp} (\textit{orange}) and Gaussian Fe-line (\textit{teal}).}
    \label{fig:fit-single}
\end{figure*}

\begin{figure*}
    \centering
    \includegraphics[width=.65\linewidth]{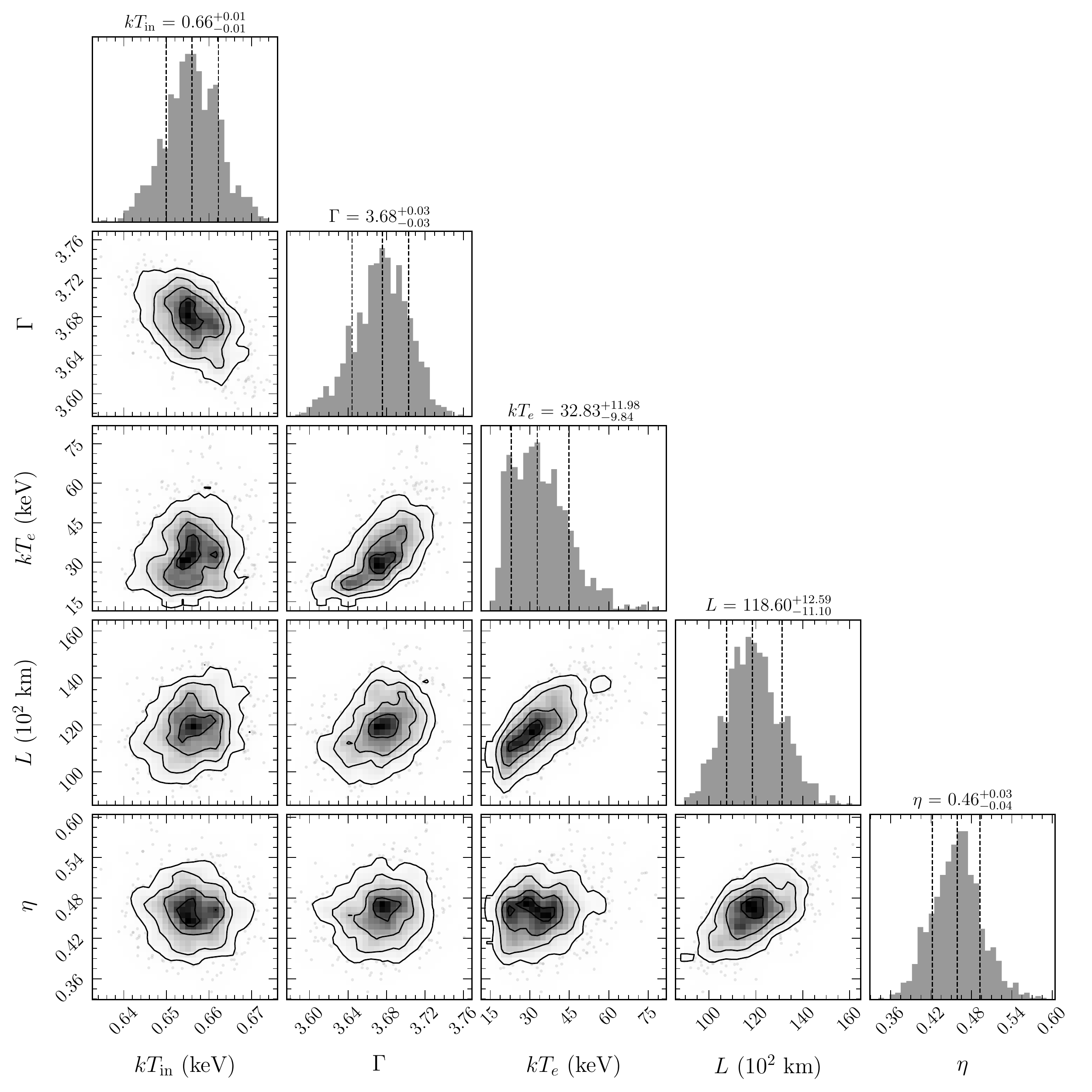}
    \caption{Corner plot of a selection of parameters of the single-corona fit to the fractional rms amplitude and phase-lag spectra of the type-B QPO, and the time-averaged energy spectrum of GX 339$-$4. The parameters, shown with their 1$\sigma$ errors in the title of each histrogram, are the temperature at the inner disc radius, $kT_\mathrm{in}$ (linked to $kT_\mathrm{bb}$ in \texttt{nthcomp} and $kT_s$ in \textsc{vkompthdk}), the power-law photon index, $\Gamma$ (linked to $\Gamma$ in \textsc{vkompthdk}), the electron temperature, $kT_e$ (linked to $kT_e$ in \textsc{vkompthdk}), the size, $L$, and the feedback parameter, $\eta$, of the corona. All the best-fitting parameters of the single-corona fit are listed in \cref{tab:fit-single}. The contours in the 2D histograms show the 68\%, 90\% and 95\% confidence levels. The dashed black lines in the diagonal histograms show the median value of the parameter and its 1$\sigma$ confidence levels.}
    \label{fig:MCMC-single}
\end{figure*}

In \cref{sec:results-model} we described the resulting coronal parameters of the single-corona fit of the \textsc{vkompthdk} model to the energy spectrum of GX 339$-$4 and the fractional rms and lag spectra of the type-B QPO. Here, in \cref{fig:fit-single} we plot the result of this joint fit of the single-corona model. In \cref{fig:MCMC-single}, we show the MCMC simulations of a sample of parameters from the fit. The panels on the diagonal of the corner plot show the posterior probabilities of the parameters of the model, with dashed black lines indicating the median value and the 1$\sigma$ confidence levels. See main text for a detailed overview and interpretation of the results.

%%%%%%%%%%%%%%%%%%%%%%%%%%%%%%%%%%%%%%%%%%%%%%%%%%

% Don't change these lines
\bsp	% typesetting comment
\label{lastpage}
\end{document}